\documentclass[sigconf, screen, authorversion,nonacm]{acmart}

\definecolor{minor}{HTML}{FF0000}
\usepackage{xcolor}

\usepackage{hyperref}
\usepackage[hyphenbreaks]{breakurl}
\usepackage{graphicx}
\usepackage{csquotes}
\usepackage{makecell}
\usepackage{subcaption}
\usepackage{svg}
\usepackage{enumitem}

\fancypagestyle{firstpage}
{
    \fancyhead[L]{Workshop paper presented at Inter.HAI'23 – the \href{https://sites.google.com/view/interhai/home}{first workshop on \textbf{Inter}disciplinary Approaches in \textbf{H}uman-\textbf{A}gent \textbf{I}nteraction}, held in conjunction with the \href{https://hai-conference.net/hai2023/}{International Conference on Human-Agent Interaction}, December 4th, 2023.}    
    \fancyhead[R]{}
}

\copyrightyear{2023} 
\acmYear{2023} 
\setcopyright{rightsretained} 
\acmConference[HAI '23]{International Conference on Human-Agent Interaction}{December 4--7, 2023}{Gothenburg, Sweden}
\acmBooktitle{International Conference on Human-Agent Interaction (HAI '23), December 4--7, 2023, Gothenburg, Sweden}\acmDOI{10.1145/3623809.3623932}
\acmISBN{979-8-4007-0824-4/23/12}

\begin{document}

\title{The Use of Multiple Conversational Agent Interlocutors in Learning}


\author{Samuel Rhys Cox}
\email{samuel.cox@u.nus.edu}
\orcid{0000-0002-4558-6610}
\affiliation{%
  \institution{National University of Singapore}
  \country{}
  }


\begin{abstract}
With growing capabilities of large language models (LLMs) comes growing affordances for human-like and context-aware conversational partners.
On from this, some recent work has investigated the use of LLMs to simulate multiple conversational partners, such as to assist users with problem solving or to simulate an environment populated entirely with LLMs.
Beyond this, we are interested in discussing and exploring the use of LLMs to simulate multiple personas to assist and augment users in educational settings that could benefit from multiple interlocutors.
We discuss prior work that uses LLMs to simulate multiple personas sharing the same environment, and discuss example scenarios where multiple conversational agent partners could be used in education.


\end{abstract}



\maketitle


\thispagestyle{firstpage}

\section{Introduction}


With advances in large language models (LLMs), conversational agents (CAs) can simulate distinct characteristics across multiple factors, such as personality and gender \cite{jiang2023personallm,safdari2023personality}.
This added affordance coupled with the growing availability and accessible nature of using LLMs could be used to create complete and believable conversational partners for users to converse with.
On from this, the use of LLMs shows promise within education \cite{yan2023practical,kasneci2023chatgpt} (such as for medical education \cite{safranek2023role} and language learning \cite{caines2023application,kasneci2023chatgpt,Ro_2023}), and LLMs could be used to provide conversational assistance and rapport building with learners.
Recent work has also investigated the use of LLMs to simulate multiple personas that could be used within the same session \cite{wang2023survey,chen2023agentverse,zhang2023exploring,park2023generative,wang2023unleashing,wu2023autogen,park2023choicemates}. For example, a user could take part in a conversation that involves multiple CAs with varying personalities, social roles, and expertise.

Building on this, an area of interest could be the use of multiple (two or more) CAs interacting with a user to aid in learning activities.
For example, CAs could act as other students or teachers when talking to the learner (user).
Multi-party conversations such as this would allow for added affordances present in group environments, such as social comparison (e.g., counterfactual comparison \cite{sun2015leaderboard}: ``\textit{You would have scored higher than Bob if you scored one more correct}''), social support (e.g., encouragement or advice giving), and the adoption of different personas \cite{khadpe2020conceptual} (e.g., personalities: relaxed, strict, friendly; or social roles).
Depending on the design and make-up of such a group, various learning theories could be harnessed such as cooperative learning, collaborative learning and peer learning.


In this workshop paper, we wish to create a discussion around the potential use of such environments where multiple CAs could interact with a user to assist in learning activities.
We will first outline some prior work that used multiple conversational agents, before highlighting some potential scenarios where the use of multiple conversational partners could be incorporated in an educational setting.

\section{Chatting Environments with Multiple Conversational Agents}




The use of LLMs to adopt diverse personas \cite{safdari2023personality,jiang2023personallm,cox2023LLM} has been harnessed both in environments that consist only of groups of LLMs communicating with each other \cite{hong2023metagpt,chen2023agentverse,li2023you,park2023generative,chan2023chateval,wang2023unleashing}, and environments where users can interact with multiple LLM-driven personas \cite{wang2023survey,zhang2023exploring,wu2023autogen,park2023choicemates,zhang2023investigating}.

In environments consisting exclusively of LLM-driven agents, agents have been shown to: behave socially when communicating, cooperate to complete tasks, and adjust their own behaviour based on the actions of other agents.
For example, \href{https://chirper.ai/}{Chirper.ai} is a social network populated entirely by LLM-driven social bots, where analysis by Li et al. \cite{li2023you} found that agents were able to exhibit individual personas and behave socially, as well as influence the behaviour of other agents;
and Park et al. \cite{park2023generative} found that a group of LLMs could simulate members of a community that completed daily activities (such as gardening and cooking), and collaborated to achieve tasks (such as planning a party for one of the agents).
Additionally, prior work has harnessed multiple LLMs as a prompting framework to solve problems by dividing work into microtasks or having agents debate amongst themselves \cite{hong2023metagpt,chen2023agentverse,chan2023chateval,wang2023unleashing}.
For example, Chan et al. \cite{chan2023chateval} found that multiple LLM agents could use debate to provide higher quality evaluations than an agent acting alone.

Environments where a user can interact with multiple CAs simultaneously have been used to simulate social environments, and assist users in problem-solving and decision-making tasks 
\cite{kitamura2002multiple,kitamura2004web,li2023leaders,park2023choicemates,jiang2023communitybots}.
For example, CommunityBots \cite{jiang2023communitybots} found that (when answering survey questions) users had higher levels of enjoyment and engagement when (sequentially) talking to three domain specific chatbots rather than talking to one chatbot.
In ChoiceMates, Park et al. \cite{park2023choicemates} compared product search tasks using web search, single LLM agent and multi-LLM agents, and found that using multiple agents helped users explore more breadth and depth of options compared to web search.





\section{Potential Scenarios for multiple CA interlocators in education}

Now that we have given an overview of some prior work that used multiple CAs, we will describe some potential scenarios where multiple CAs could be used in education.

\subsection{Differing Social Roles in Conversations}

Prior work has investigated the effect of varying the social role of an agent \cite{howley2014effects,khadpe2020conceptual,shirouzu2013effects}, such as comparing a social robot in a classroom helping students as either a ``co-solver'' or a ``knower'' \cite{shirouzu2013effects}.
Building on this, by adopting multiple conversational agents into a chatting session with a learner, we could take advantage of the affordance for diverse social roles.
For example, in a scenario where a learner is required to explain and discuss a complex scientific concept, the learner could be conversing with multiple agents, such as: a sceptic agent who could challenge a learner's responses to ensure a robust subject understanding that withstands scrutiny; an encouraging agent that could provide positive reinforcement and highlight a learner's progress; and a mentor agent that could offer additional guidance and insights or connect concepts to real-world examples.

\subsection{Conversational Partners with Different Cultural Background and Knowledge}

Multiple conversational partners could also prove beneficial in learning fields where diverse perspectives would aid learners in becoming more well-rounded and aware of nuance and variation.
For example, within language learning the same language could have slight variations due to the different cultural and linguistic backgrounds.
For example, in Singapore people may speak a colloquial form of Singaporean Mandarin that combines Mandarin, English and Malay vocabulary; or they may speak Standard Mandarin depending on their situation and interlocutor. By allowing for group conversations that incorporate multiple different linguistic variations, learners could be aware of different contexts and vocabulary that could be used depending on the context of a given situation.

\subsection{Virtual Environments with LLM Interlocutors}

Drawing comparison to Park et al.'s work where a virtual AI Town was populated entirely by LLMs that could interact with each other and the environment \cite{park2023generative}, virtual environments could be developed that are populated by embodied LLM-driven CAs to allow for an immersion-learning environment. 
Human users could then observe interactions between agents (that could act as exemplar interactions in training exercises such as conflict resolution \cite{shaikh2023rehearsal} or language learning), or users themselves could interact with agents (to gain experience in practical interaction skills that the user may be pursuing).

However, depending on the level of LLM control and capabilities, it should be ensured that agents do not mislead or discomfort users, such as via hallucinations, contradicting information disclosure from LLMs, or inappropriate utterances that make users uncomfortable (see \cite{cox2023gaming} for guidelines to design conversational interactions with LLM-driven computer-controlled characters).

\section{Discussion and Conclusion}


In conclusion, we have discussed recent work that uses LLMs to simulate multiple conversational partners that would have the potential to aid learners.
While the scope of such interactions is very broad (allowing for the adoption of multiple pedagogical techniques, or interactions paradigms that rely on the presence of three or more interlocutors), we hope to raise discussion.

Attention should also be drawn to well-known issues surrounding LLMs such as potential for hallucinations (that could prove counter-productive in an educational environment if the learner is provided with mistruths), and bias or stereotypes in LLM responses that could prove harmful to learners.

The utilisation of multiple CAs (in not just education, but any application domain) adds additional layers of complexity, whereby agents need to be designed such that the CA personas (or teach methods) harmonise in a manner proving helpful and appropriate to learners. Additionally, controls are needed to ensure CAs do not provide contradictory information.
Interaction dynamics between the agents themselves also need to be crafted carefully to ensure that they simulate productive and appropriate human interactions.


Is it also important to discuss potential of over-reliance on such systems. Should systems be developed that promote independent thought, and avoid environments where learners become passive.
On from this, additional ethical concerns should be considered, such as handling of personal data, potential for bias in agent responses (that could prove either uncomfortable and inappropriate to learners, or could introduce biased utterances as exemplar interactions to learners depending on the learning environment), and the impact of environments and interactions on learner's well-being.


\bibliographystyle{ACM-Reference-Format}
\bibliography{sample-base}


\begin{thebibliography}{29}


\ifx \showCODEN    \undefined \def \showCODEN     #1{\unskip}     \fi
\ifx \showDOI      \undefined \def \showDOI       #1{#1}\fi
\ifx \showISBNx    \undefined \def \showISBNx     #1{\unskip}     \fi
\ifx \showISBNxiii \undefined \def \showISBNxiii  #1{\unskip}     \fi
\ifx \showISSN     \undefined \def \showISSN      #1{\unskip}     \fi
\ifx \showLCCN     \undefined \def \showLCCN      #1{\unskip}     \fi
\ifx \shownote     \undefined \def \shownote      #1{#1}          \fi
\ifx \showarticletitle \undefined \def \showarticletitle #1{#1}   \fi
\ifx \showURL      \undefined \def \showURL       {\relax}        \fi
\providecommand\bibfield[2]{#2}
\providecommand\bibinfo[2]{#2}
\providecommand\natexlab[1]{#1}
\providecommand\showeprint[2][]{arXiv:#2}

\bibitem[\protect\citeauthoryear{Caines, Benedetto, Taslimipoor, Davis, Gao, Andersen, Yuan, Elliott, Moore, Bryant, et~al\mbox{.}}{Caines et~al\mbox{.}}{2023}]%
        {caines2023application}
\bibfield{author}{\bibinfo{person}{Andrew Caines}, \bibinfo{person}{Luca Benedetto}, \bibinfo{person}{Shiva Taslimipoor}, \bibinfo{person}{Christopher Davis}, \bibinfo{person}{Yuan Gao}, \bibinfo{person}{Oeistein Andersen}, \bibinfo{person}{Zheng Yuan}, \bibinfo{person}{Mark Elliott}, \bibinfo{person}{Russell Moore}, \bibinfo{person}{Christopher Bryant}, {et~al\mbox{.}}} \bibinfo{year}{2023}\natexlab{}.
\newblock \showarticletitle{On the application of Large Language Models for language teaching and assessment technology}.
\newblock \bibinfo{journal}{\emph{arXiv preprint arXiv:2307.08393}} (\bibinfo{year}{2023}).
\newblock


\bibitem[\protect\citeauthoryear{Chan, Chen, Su, Yu, Xue, Zhang, Fu, and Liu}{Chan et~al\mbox{.}}{2023}]%
        {chan2023chateval}
\bibfield{author}{\bibinfo{person}{Chi-Min Chan}, \bibinfo{person}{Weize Chen}, \bibinfo{person}{Yusheng Su}, \bibinfo{person}{Jianxuan Yu}, \bibinfo{person}{Wei Xue}, \bibinfo{person}{Shanghang Zhang}, \bibinfo{person}{Jie Fu}, {and} \bibinfo{person}{Zhiyuan Liu}.} \bibinfo{year}{2023}\natexlab{}.
\newblock \showarticletitle{Chateval: Towards better llm-based evaluators through multi-agent debate}.
\newblock \bibinfo{journal}{\emph{arXiv preprint arXiv:2308.07201}} (\bibinfo{year}{2023}).
\newblock


\bibitem[\protect\citeauthoryear{Chen, Su, Zuo, Yang, Yuan, Qian, Chan, Qin, Lu, Xie, et~al\mbox{.}}{Chen et~al\mbox{.}}{2023}]%
        {chen2023agentverse}
\bibfield{author}{\bibinfo{person}{Weize Chen}, \bibinfo{person}{Yusheng Su}, \bibinfo{person}{Jingwei Zuo}, \bibinfo{person}{Cheng Yang}, \bibinfo{person}{Chenfei Yuan}, \bibinfo{person}{Chen Qian}, \bibinfo{person}{Chi-Min Chan}, \bibinfo{person}{Yujia Qin}, \bibinfo{person}{Yaxi Lu}, \bibinfo{person}{Ruobing Xie}, {et~al\mbox{.}}} \bibinfo{year}{2023}\natexlab{}.
\newblock \showarticletitle{Agentverse: Facilitating multi-agent collaboration and exploring emergent behaviors in agents}.
\newblock \bibinfo{journal}{\emph{arXiv preprint arXiv:2308.10848}} (\bibinfo{year}{2023}).
\newblock


\bibitem[\protect\citeauthoryear{Cox, Abdul, and Ooi}{Cox et~al\mbox{.}}{2023}]%
        {cox2023LLM}
\bibfield{author}{\bibinfo{person}{Samuel~Rhys Cox}, \bibinfo{person}{Ashraf Abdul}, {and} \bibinfo{person}{Wei~Tsang Ooi}.} \bibinfo{year}{2023}\natexlab{}.
\newblock \showarticletitle{Prompting a Large Language Model to Generate Diverse Motivational Messages: A Comparison with Human-Written Messages}. In \bibinfo{booktitle}{\emph{Proceedings of the 11th International Conference on Human-Agent Interaction}}.
\newblock
\urldef\tempurl%
\url{https://doi.org/10.48550/arXiv.2308.13479}
\showURL{%
\tempurl}


\bibitem[\protect\citeauthoryear{Cox and Ooi}{Cox and Ooi}{2023}]%
        {cox2023gaming}
\bibfield{author}{\bibinfo{person}{Samuel~Rhys Cox} {and} \bibinfo{person}{Wei~Tsang Ooi}.} \bibinfo{year}{2023}\natexlab{}.
\newblock \showarticletitle{Conversational Interactions with NPCs in LLM-Driven Gaming: Guidelines from a Content Analysis of Player Feedback}. In \bibinfo{booktitle}{\emph{the 7th International Workshop on Chatbot Research and Design}}. Springer.
\newblock
\urldef\tempurl%
\url{https://conversations2023.files.wordpress.com/2023/11/conversations_2023_preprint_36_cox.pdf}
\showURL{%
\tempurl}


\bibitem[\protect\citeauthoryear{Hong, Zheng, Chen, Cheng, Zhang, Wang, Yau, Lin, Zhou, Ran, et~al\mbox{.}}{Hong et~al\mbox{.}}{2023}]%
        {hong2023metagpt}
\bibfield{author}{\bibinfo{person}{Sirui Hong}, \bibinfo{person}{Xiawu Zheng}, \bibinfo{person}{Jonathan Chen}, \bibinfo{person}{Yuheng Cheng}, \bibinfo{person}{Ceyao Zhang}, \bibinfo{person}{Zili Wang}, \bibinfo{person}{Steven Ka~Shing Yau}, \bibinfo{person}{Zijuan Lin}, \bibinfo{person}{Liyang Zhou}, \bibinfo{person}{Chenyu Ran}, {et~al\mbox{.}}} \bibinfo{year}{2023}\natexlab{}.
\newblock \showarticletitle{Metagpt: Meta programming for multi-agent collaborative framework}.
\newblock \bibinfo{journal}{\emph{arXiv preprint arXiv:2308.00352}} (\bibinfo{year}{2023}).
\newblock


\bibitem[\protect\citeauthoryear{Howley, Kanda, Hayashi, and Ros{\'e}}{Howley et~al\mbox{.}}{2014}]%
        {howley2014effects}
\bibfield{author}{\bibinfo{person}{Iris Howley}, \bibinfo{person}{Takayuki Kanda}, \bibinfo{person}{Kotaro Hayashi}, {and} \bibinfo{person}{Carolyn Ros{\'e}}.} \bibinfo{year}{2014}\natexlab{}.
\newblock \showarticletitle{Effects of social presence and social role on help-seeking and learning}. In \bibinfo{booktitle}{\emph{Proceedings of the 2014 ACM/IEEE international conference on Human-robot interaction}}. \bibinfo{pages}{415--422}.
\newblock


\bibitem[\protect\citeauthoryear{Jiang, Zhang, Cao, Kabbara, and Roy}{Jiang et~al\mbox{.}}{2023b}]%
        {jiang2023personallm}
\bibfield{author}{\bibinfo{person}{Hang Jiang}, \bibinfo{person}{Xiajie Zhang}, \bibinfo{person}{Xubo Cao}, \bibinfo{person}{Jad Kabbara}, {and} \bibinfo{person}{Deb Roy}.} \bibinfo{year}{2023}\natexlab{b}.
\newblock \showarticletitle{{PersonaLLM: Investigating the Ability of GPT-3.5 to Express Personality Traits and Gender Differences}}.
\newblock \bibinfo{journal}{\emph{Proceedings of the 9th International Conference on Computational Social Science}}.
\newblock
\urldef\tempurl%
\url{https://doi.org/10.48550/arXiv.2305.02547}
\showURL{%
\tempurl}


\bibitem[\protect\citeauthoryear{Jiang, Rashik, Panchal, Jasim, Sarvghad, Riahi, DeWitt, Thurber, and Mahyar}{Jiang et~al\mbox{.}}{2023a}]%
        {jiang2023communitybots}
\bibfield{author}{\bibinfo{person}{Zhiqiu Jiang}, \bibinfo{person}{Mashrur Rashik}, \bibinfo{person}{Kunjal Panchal}, \bibinfo{person}{Mahmood Jasim}, \bibinfo{person}{Ali Sarvghad}, \bibinfo{person}{Pari Riahi}, \bibinfo{person}{Erica DeWitt}, \bibinfo{person}{Fey Thurber}, {and} \bibinfo{person}{Narges Mahyar}.} \bibinfo{year}{2023}\natexlab{a}.
\newblock \showarticletitle{CommunityBots: Creating and Evaluating A Multi-Agent Chatbot Platform for Public Input Elicitation}.
\newblock \bibinfo{journal}{\emph{Proceedings of the ACM on Human-Computer Interaction}} \bibinfo{volume}{7}, \bibinfo{number}{CSCW1} (\bibinfo{year}{2023}), \bibinfo{pages}{1--32}.
\newblock


\bibitem[\protect\citeauthoryear{Kasneci, Se{\ss}ler, K{\"u}chemann, Bannert, Dementieva, Fischer, Gasser, Groh, G{\"u}nnemann, H{\"u}llermeier, et~al\mbox{.}}{Kasneci et~al\mbox{.}}{2023}]%
        {kasneci2023chatgpt}
\bibfield{author}{\bibinfo{person}{Enkelejda Kasneci}, \bibinfo{person}{Kathrin Se{\ss}ler}, \bibinfo{person}{Stefan K{\"u}chemann}, \bibinfo{person}{Maria Bannert}, \bibinfo{person}{Daryna Dementieva}, \bibinfo{person}{Frank Fischer}, \bibinfo{person}{Urs Gasser}, \bibinfo{person}{Georg Groh}, \bibinfo{person}{Stephan G{\"u}nnemann}, \bibinfo{person}{Eyke H{\"u}llermeier}, {et~al\mbox{.}}} \bibinfo{year}{2023}\natexlab{}.
\newblock \showarticletitle{ChatGPT for good? On opportunities and challenges of large language models for education}.
\newblock \bibinfo{journal}{\emph{Learning and individual differences}}  \bibinfo{volume}{103} (\bibinfo{year}{2023}), \bibinfo{pages}{102274}.
\newblock


\bibitem[\protect\citeauthoryear{Khadpe, Krishna, Fei-Fei, Hancock, and Bernstein}{Khadpe et~al\mbox{.}}{2020}]%
        {khadpe2020conceptual}
\bibfield{author}{\bibinfo{person}{Pranav Khadpe}, \bibinfo{person}{Ranjay Krishna}, \bibinfo{person}{Li Fei-Fei}, \bibinfo{person}{Jeffrey~T Hancock}, {and} \bibinfo{person}{Michael~S Bernstein}.} \bibinfo{year}{2020}\natexlab{}.
\newblock \showarticletitle{Conceptual metaphors impact perceptions of human-AI collaboration}.
\newblock \bibinfo{journal}{\emph{Proceedings of the ACM on Human-Computer Interaction}} \bibinfo{volume}{4}, \bibinfo{number}{CSCW2} (\bibinfo{year}{2020}), \bibinfo{pages}{1--26}.
\newblock


\bibitem[\protect\citeauthoryear{Kitamura}{Kitamura}{2004}]%
        {kitamura2004web}
\bibfield{author}{\bibinfo{person}{Yasuhiko Kitamura}.} \bibinfo{year}{2004}\natexlab{}.
\newblock \showarticletitle{Web information integration using multiple character agents}.
\newblock \bibinfo{journal}{\emph{Life-Like Characters: Tools, Affective Functions, and Applications}} (\bibinfo{year}{2004}), \bibinfo{pages}{295--315}.
\newblock


\bibitem[\protect\citeauthoryear{Kitamura, Tsujimoto, Yamada, and Yamamoto}{Kitamura et~al\mbox{.}}{2002}]%
        {kitamura2002multiple}
\bibfield{author}{\bibinfo{person}{Yasuhiko Kitamura}, \bibinfo{person}{Hideki Tsujimoto}, \bibinfo{person}{Teruhiro Yamada}, {and} \bibinfo{person}{Taizo Yamamoto}.} \bibinfo{year}{2002}\natexlab{}.
\newblock \showarticletitle{Multiple character-agents interface: An information integration platform where multiple agents and human user collaborate}. In \bibinfo{booktitle}{\emph{Proceedings of the first international joint conference on Autonomous agents and multiagent systems: part 2}}. \bibinfo{pages}{790--791}.
\newblock


\bibitem[\protect\citeauthoryear{Li, Yang, and Zhao}{Li et~al\mbox{.}}{2023a}]%
        {li2023you}
\bibfield{author}{\bibinfo{person}{Siyu Li}, \bibinfo{person}{Jin Yang}, {and} \bibinfo{person}{Kui Zhao}.} \bibinfo{year}{2023}\natexlab{a}.
\newblock \showarticletitle{Are you in a masquerade? Exploring the behavior and impact of large language model driven social bots in online social networks}.
\newblock \bibinfo{journal}{\emph{arXiv preprint arXiv:2307.10337}} (\bibinfo{year}{2023}).
\newblock
\urldef\tempurl%
\url{https://doi.org/10.48550/arXiv.2307.10337}
\showURL{%
\tempurl}


\bibitem[\protect\citeauthoryear{Li, Yuan, Zhao, and Yang}{Li et~al\mbox{.}}{2023b}]%
        {li2023leaders}
\bibfield{author}{\bibinfo{person}{Shuo Li}, \bibinfo{person}{Xiang Yuan}, \bibinfo{person}{Xinyuan Zhao}, {and} \bibinfo{person}{Shirao Yang}.} \bibinfo{year}{2023}\natexlab{b}.
\newblock \showarticletitle{Leaders or Team-Mates: Exploring the Role-Based Relationship Between Multiple Intelligent Agents in Driving Scenarios: Research on the Role-Based Relationship Between Multiple Intelligent Agents in Driving Scenarios}. In \bibinfo{booktitle}{\emph{International Conference on Human-Computer Interaction}}. Springer, \bibinfo{pages}{144--165}.
\newblock


\bibitem[\protect\citeauthoryear{Park, Min, Ma, and Kim}{Park et~al\mbox{.}}{2023a}]%
        {park2023choicemates}
\bibfield{author}{\bibinfo{person}{Jeongeon Park}, \bibinfo{person}{Bryan Min}, \bibinfo{person}{Xiaojuan Ma}, {and} \bibinfo{person}{Juho Kim}.} \bibinfo{year}{2023}\natexlab{a}.
\newblock \bibinfo{title}{ChoiceMates: Supporting Unfamiliar Online Decision-Making with Multi-Agent Conversational Interactions}.
\newblock
\newblock
\showeprint[arxiv]{2310.01331}~[cs.HC]


\bibitem[\protect\citeauthoryear{Park, O'Brien, Cai, Morris, Liang, and Bernstein}{Park et~al\mbox{.}}{2023b}]%
        {park2023generative}
\bibfield{author}{\bibinfo{person}{Joon~Sung Park}, \bibinfo{person}{Joseph~C O'Brien}, \bibinfo{person}{Carrie~J Cai}, \bibinfo{person}{Meredith~Ringel Morris}, \bibinfo{person}{Percy Liang}, {and} \bibinfo{person}{Michael~S Bernstein}.} \bibinfo{year}{2023}\natexlab{b}.
\newblock \showarticletitle{Generative agents: Interactive simulacra of human behavior}.
\newblock \bibinfo{journal}{\emph{ACM Designing Intelligent Systems (DIS)}} (\bibinfo{year}{2023}).
\newblock


\bibitem[\protect\citeauthoryear{Ro}{Ro}{2023}]%
        {Ro_2023}
\bibfield{author}{\bibinfo{person}{Christine Ro}.} \bibinfo{year}{2023}\natexlab{}.
\newblock \bibinfo{title}{Students switch to AI to learn languages}.
\newblock
\newblock
\urldef\tempurl%
\url{https://www.bbc.com/news/business-65849104}
\showURL{%
\tempurl}


\bibitem[\protect\citeauthoryear{Safdari, Serapio-Garc{\'\i}a, Crepy, Fitz, Romero, Sun, Abdulhai, Faust, and Matari{\'c}}{Safdari et~al\mbox{.}}{2023}]%
        {safdari2023personality}
\bibfield{author}{\bibinfo{person}{Mustafa Safdari}, \bibinfo{person}{Greg Serapio-Garc{\'\i}a}, \bibinfo{person}{Cl{\'e}ment Crepy}, \bibinfo{person}{Stephen Fitz}, \bibinfo{person}{Peter Romero}, \bibinfo{person}{Luning Sun}, \bibinfo{person}{Marwa Abdulhai}, \bibinfo{person}{Aleksandra Faust}, {and} \bibinfo{person}{Maja Matari{\'c}}.} \bibinfo{year}{2023}\natexlab{}.
\newblock \showarticletitle{Personality traits in large language models}.
\newblock \bibinfo{journal}{\emph{arXiv preprint arXiv:2307.00184}} (\bibinfo{year}{2023}).
\newblock
\urldef\tempurl%
\url{https://doi.org/10.48550/arXiv.2307.00184}
\showURL{%
\tempurl}


\bibitem[\protect\citeauthoryear{Safranek, Sidamon-Eristoff, Gilson, and Chartash}{Safranek et~al\mbox{.}}{2023}]%
        {safranek2023role}
\bibfield{author}{\bibinfo{person}{Conrad~W Safranek}, \bibinfo{person}{Anne~Elizabeth Sidamon-Eristoff}, \bibinfo{person}{Aidan Gilson}, {and} \bibinfo{person}{David Chartash}.} \bibinfo{year}{2023}\natexlab{}.
\newblock \bibinfo{title}{The Role of Large Language Models in Medical Education: Applications and Implications}.
\newblock , \bibinfo{numpages}{e50945}~pages.
\newblock


\bibitem[\protect\citeauthoryear{Shaikh, Chai, Gelfand, Yang, and Bernstein}{Shaikh et~al\mbox{.}}{2023}]%
        {shaikh2023rehearsal}
\bibfield{author}{\bibinfo{person}{Omar Shaikh}, \bibinfo{person}{Valentino Chai}, \bibinfo{person}{Michele~J Gelfand}, \bibinfo{person}{Diyi Yang}, {and} \bibinfo{person}{Michael~S Bernstein}.} \bibinfo{year}{2023}\natexlab{}.
\newblock \showarticletitle{Rehearsal: Simulating Conflict to Teach Conflict Resolution}.
\newblock \bibinfo{journal}{\emph{arXiv preprint arXiv:2309.12309}} (\bibinfo{year}{2023}).
\newblock


\bibitem[\protect\citeauthoryear{Shirouzu and Miyake}{Shirouzu and Miyake}{2013}]%
        {shirouzu2013effects}
\bibfield{author}{\bibinfo{person}{Hajime Shirouzu} {and} \bibinfo{person}{Naomi Miyake}.} \bibinfo{year}{2013}\natexlab{}.
\newblock \showarticletitle{Effects of Robots' Revoicing on Preparation for Future Learning}.
\newblock  (\bibinfo{year}{2013}).
\newblock


\bibitem[\protect\citeauthoryear{Sun, Jones, Traca, and Bos}{Sun et~al\mbox{.}}{2015}]%
        {sun2015leaderboard}
\bibfield{author}{\bibinfo{person}{Emily Sun}, \bibinfo{person}{Brooke Jones}, \bibinfo{person}{Stefano Traca}, {and} \bibinfo{person}{Maarten~W Bos}.} \bibinfo{year}{2015}\natexlab{}.
\newblock \showarticletitle{Leaderboard position psychology: counterfactual thinking}. In \bibinfo{booktitle}{\emph{Proceedings of the 33rd Annual ACM Conference Extended Abstracts on Human Factors in Computing Systems}}. \bibinfo{pages}{1217--1222}.
\newblock


\bibitem[\protect\citeauthoryear{Wang, Ma, Feng, Zhang, Yang, Zhang, Chen, Tang, Chen, Lin, et~al\mbox{.}}{Wang et~al\mbox{.}}{2023a}]%
        {wang2023survey}
\bibfield{author}{\bibinfo{person}{Lei Wang}, \bibinfo{person}{Chen Ma}, \bibinfo{person}{Xueyang Feng}, \bibinfo{person}{Zeyu Zhang}, \bibinfo{person}{Hao Yang}, \bibinfo{person}{Jingsen Zhang}, \bibinfo{person}{Zhiyuan Chen}, \bibinfo{person}{Jiakai Tang}, \bibinfo{person}{Xu Chen}, \bibinfo{person}{Yankai Lin}, {et~al\mbox{.}}} \bibinfo{year}{2023}\natexlab{a}.
\newblock \showarticletitle{A survey on large language model based autonomous agents}.
\newblock \bibinfo{journal}{\emph{arXiv preprint arXiv:2308.11432}} (\bibinfo{year}{2023}).
\newblock


\bibitem[\protect\citeauthoryear{Wang, Mao, Wu, Ge, Wei, and Ji}{Wang et~al\mbox{.}}{2023b}]%
        {wang2023unleashing}
\bibfield{author}{\bibinfo{person}{Zhenhailong Wang}, \bibinfo{person}{Shaoguang Mao}, \bibinfo{person}{Wenshan Wu}, \bibinfo{person}{Tao Ge}, \bibinfo{person}{Furu Wei}, {and} \bibinfo{person}{Heng Ji}.} \bibinfo{year}{2023}\natexlab{b}.
\newblock \showarticletitle{Unleashing cognitive synergy in large language models: A task-solving agent through multi-persona self-collaboration}.
\newblock \bibinfo{journal}{\emph{arXiv preprint arXiv:2307.05300}} (\bibinfo{year}{2023}).
\newblock


\bibitem[\protect\citeauthoryear{Wu, Bansal, Zhang, Wu, Zhang, Zhu, Li, Jiang, Zhang, and Wang}{Wu et~al\mbox{.}}{2023}]%
        {wu2023autogen}
\bibfield{author}{\bibinfo{person}{Qingyun Wu}, \bibinfo{person}{Gagan Bansal}, \bibinfo{person}{Jieyu Zhang}, \bibinfo{person}{Yiran Wu}, \bibinfo{person}{Shaokun Zhang}, \bibinfo{person}{Erkang Zhu}, \bibinfo{person}{Beibin Li}, \bibinfo{person}{Li Jiang}, \bibinfo{person}{Xiaoyun Zhang}, {and} \bibinfo{person}{Chi Wang}.} \bibinfo{year}{2023}\natexlab{}.
\newblock \showarticletitle{Autogen: Enabling next-gen LLM applications via multi-agent conversation framework}.
\newblock \bibinfo{journal}{\emph{arXiv preprint arXiv:2308.08155}} (\bibinfo{year}{2023}).
\newblock


\bibitem[\protect\citeauthoryear{Yan, Sha, Zhao, Li, Martinez-Maldonado, Chen, Li, Jin, and Ga{\v{s}}evi{\'c}}{Yan et~al\mbox{.}}{2023}]%
        {yan2023practical}
\bibfield{author}{\bibinfo{person}{Lixiang Yan}, \bibinfo{person}{Lele Sha}, \bibinfo{person}{Linxuan Zhao}, \bibinfo{person}{Yuheng Li}, \bibinfo{person}{Roberto Martinez-Maldonado}, \bibinfo{person}{Guanliang Chen}, \bibinfo{person}{Xinyu Li}, \bibinfo{person}{Yueqiao Jin}, {and} \bibinfo{person}{Dragan Ga{\v{s}}evi{\'c}}.} \bibinfo{year}{2023}\natexlab{}.
\newblock \showarticletitle{Practical and ethical challenges of large language models in education: A systematic scoping review}.
\newblock \bibinfo{journal}{\emph{British Journal of Educational Technology}} (\bibinfo{year}{2023}).
\newblock


\bibitem[\protect\citeauthoryear{Zhang, Xu, and Deng}{Zhang et~al\mbox{.}}{2023b}]%
        {zhang2023exploring}
\bibfield{author}{\bibinfo{person}{Jintian Zhang}, \bibinfo{person}{Xin Xu}, {and} \bibinfo{person}{Shumin Deng}.} \bibinfo{year}{2023}\natexlab{b}.
\newblock \bibinfo{title}{Exploring Collaboration Mechanisms for LLM Agents: A Social Psychology View}.
\newblock
\newblock
\showeprint[arxiv]{2310.02124}~[cs.CL]
\urldef\tempurl%
\url{https://doi.org/10.48550/arXiv.2310.02124}
\showURL{%
\tempurl}


\bibitem[\protect\citeauthoryear{Zhang, Duan, Flathmann, McNeese, Freeman, and Williams}{Zhang et~al\mbox{.}}{2023a}]%
        {zhang2023investigating}
\bibfield{author}{\bibinfo{person}{Rui Zhang}, \bibinfo{person}{Wen Duan}, \bibinfo{person}{Christopher Flathmann}, \bibinfo{person}{Nathan McNeese}, \bibinfo{person}{Guo Freeman}, {and} \bibinfo{person}{Alyssa Williams}.} \bibinfo{year}{2023}\natexlab{a}.
\newblock \showarticletitle{Investigating AI Teammate Communication Strategies and Their Impact in Human-AI Teams for Effective Teamwork}.
\newblock \bibinfo{journal}{\emph{Proceedings of the ACM on Human-Computer Interaction}} \bibinfo{volume}{7}, \bibinfo{number}{CSCW2} (\bibinfo{year}{2023}), \bibinfo{pages}{1--31}.
\newblock


\end{thebibliography}

\end{document}